\documentclass[12pt]{article}

\usepackage{amsmath,amssymb,amsbsy,amsfonts,amsthm,latexsym,
                        amsopn,amstext,amsxtra,euscript,amscd,color}
\usepackage{color,xcolor}
\usepackage{latexsym}
\usepackage{cite}
\usepackage{amsfonts}
\usepackage{amsfonts,mathrsfs}
\usepackage{graphicx}
\usepackage{psfrag}
\usepackage{subfigure}
\usepackage{url}
\usepackage{stfloats}
\usepackage{amsmath}

\newtheorem{theorem}{Theorem}
\newtheorem{lemma}{Lemma}

\newcommand{\quash}[1]{}

\begin{document}

\title{Linear complexity of quaternary sequences over $\mathbb{Z}_4$ based on Ding-Helleseth generalized cyclotomic classes}

\author{Xina Zhang$^{1}$, Xiaoni Du$^{1}$, Chenhuang Wu$^{2}$\\
1. College of Mathematics and Statistics, Northwest
Normal University, \\
Lanzhou, Gansu 730070, P.R. China\\
ymLdxn@126.com\\
2. School of Mathematics, Putian University, \\ Putian, Fujian
351100, P.R. China
}

\maketitle

\begin{abstract}

A family of quaternary sequences over $\mathbb{Z}_4$ is defined based on the  Ding-Helleseth generalized cyclotomic classes modulo $pq$  for two distinct odd primes $p$ and $q$.   The linear complexity is determined by computing the defining polynomial of the sequences, which is in fact connected with the discrete Fourier transform of the  sequences. The results show that the sequences possess large linear complexity and  are  ``good" sequences  from the viewpoint of cryptography.

\end{abstract}

\section{Introduction}

Pseudo-random sequences with sound  pseudo-randomness  properties have been widely used in modern communication systems and cryptography
\cite{Golomb-book, GG}. Cyclotomic and generalized cyclotomic sequences over finite fields are important pseudorandom sequences in stream ciphers due to their sound  pseudo-random cryptographic properties and large linear complexity, such as  Legendre sequences, Jacobi sequences, etc. \cite{DGS,DGS09,DGSY,D97}.
We should mention that most of the generalized cyclotomic sequences are defined over finite fields $\mathbb{F}_2$ or $\mathbb{F}_4$ or $\mathbb{F}_r$ with $r$ an odd prime( see \cite{DGS,DGS09,DGSY,D97,DU-Chen-cje,LWZH}, for example). For the quaternary sequences over ring $\mathbb{Z}_4$, most of the studies are  focused on the analysis of their autocorrelation\cite{KJKN, US96, YT} and the analysis of the linear complexity \cite{EI14}  are rare,  especially for the  generalized cyclotomic sequences sequences over ring $\mathbb{Z}_4$.

Very recently, as a generalization  of Jacobi sequences, a family of quaternary sequences over $\mathbb{Z}_4$ with  period $pq$  was proposed by Edemskiy\cite{EI13} and Chen  \cite{Chen-Z4}, respectively. They determined the linear complexity with different methods. In this paper,  we will propose a similar kind of  quaternary sequences over $\mathbb{Z}_4$ using the  Ding-Helleseth generalized cyclotomic classes \cite{D98}.

For cryptographic applications, the  \emph{linear complexity} $L((s_u))$ of
a $N$ period sequence $(s_u)$  is an important merit factor \cite{CDR,GAO-HU-LI,GG,Reeds-Sloane}. It
may be defined as the length of the shortest linear feedback shift
register  which generates the sequence. The feedback
function of this shift register can be deduced from the knowledge of
just $2L((s_u))$ consecutive digits of the sequence. Thus, it is reasonable
to suggest that ``good" sequences have $L((s_u)) > N / 2$  from the viewpoint of cryptography
 \cite{CDR,Reeds-Sloane}.

Let $p$ and $q$ be two distinct primes with $\gcd(p-1,q-1)=4$ and
$e=(p-1)(q-1)/4$. By the Chinese Remainder Theorem there exists a
common primitive root $g$ of both $p$ and $q$,  and the multiplicative order of $g$
modulo $pq$ is $e$.  There also exists an
integer $h$ satisfying$
h\equiv g\pmod{p}$ and  $h\equiv 1\pmod {q}.$ Define
$$
D_i=\{g^{4s+i}h^j \pmod {pq} : 0\leq s < e/4,~0\le j< 4\},~~ 0\le i<4
$$
and  thus, the multiplication subgroup of the residue ring $\mathbb{Z}_{pq}$ is
$
\mathbb{Z}_{pq}^*=\bigcup\limits_{i=0}^3 D_i.
$
We note that $h^4\in D_0$, since otherwise, we write $h^4\equiv g^{4s+i}h^j \pmod {pq}$ for some $0\le s<e/4$ and $1\le i<4$ and
get $g^{e-(4s+i)}h^{4-j}=1 \in D_i$, a contradiction.

Let $
P=\{p,2p,\ldots,(q-1)p\}, ~ Q=\{q,2q,\ldots,(p-1)q\}, ~R=\{0\}.
$
Then a quaternary sequences $(e_u)$  over $\mathbb{Z}_4$ of length $pq$  are defined by
\begin{equation}\label{quaternary}
e_u=\left\{
\begin{array}{ll}
2, & \mathrm{if}\,\ u\pmod {pq}\in Q\cup R,\\
0, & \mathrm{if}\,\ u\pmod {pq}\in P,\\
i, & \mathrm{if}\,\ u\pmod {pq}\in D_i, i=0,1,2,3.
\end{array}
\right.
\end{equation}

In Section 2, we will compute the  defining polynomial of  $(e_u)$ in (\ref{quaternary}) first and then determine exact values of the linear complexity.  The  defining polynomial of $(e_u)$ can also help us to give the trace representation of  $(e_u)$, which we mention in Section 3. In the rest of the paper, we always suppose that the subscript of  $D$ is performed modulo $4$.

\section{Main Results  and Proof}

First we introduce the  defining pairs  of the sequence  $(s_u)$ over $\mathbb{Z}_4$ with odd period $T$.  The group of units of Galois ring $GR(4,4^r)$ of characteristic 4 with  $4^r$ many elements, is $GR^*(4,4^r)=G_1\times G_2,$ where $G_1$ is a cyclic group of order $2^r-1$ and $G_2$ is a group of order $2^r$. Note that
any $\alpha\in GR(4,4^r)$  can be uniquely represented as
\begin{equation}\label{elem}
\alpha := \alpha_1+2\alpha_2, ~~~ \alpha_1,\alpha_2\in\mathcal{T}.
\end{equation}
where $\mathcal{T}=\{0\}\cup G_1$. See \cite[Ch. 14]{W} for details on the theory of Galois rings.

Let  $T|(2^r-1)$ and $\alpha\in GR(4,4^r)$ be a primitive $T$-th root
of unity. By \cite{US}, one can see  that
$
s_u=\sum\limits_{0\le i<T}\rho_i\alpha^{iu},
$
where $
\rho_i=\sum\limits_{0\le u<T}s_u\alpha^{-iu}~(0\le i<T)
$ is
the (discrete) Fourier transform (DFT) of $(s_u)$. We call $
s_u=G(\alpha^u), u\ge 0$ with $G(X)= \sum\limits_{0\le i<T}\rho_iX^i \in  GR(4,4^r)[X]$ the \emph{defining polynomial} of $(s_u)$ corresponding to $\alpha$  \cite{DGS,DGS09,DGSY} and $(G(X),\alpha)$  a \emph{defining pair} of $(s_u)$.

Define
$$
D_i(X)= \sum\limits_{u\in D_i}X^u \in
\mathbb{Z}_4[X]
$$
for $i=0,1,2,3$.

From our construction, one can see that $p$ and $q$ satisfy one of $q\equiv 1 \pmod 8$ and $p\equiv 5 \pmod 8$ or $q\equiv 5 \pmod 8$ and  $p\equiv 1 \pmod 4$ since $\gcd(p-1,q-1)=4$. Now we present the defining polynomial of the sequences  as following.

%\begin{theorem}\label{def-poly-caseone}
%With the notations  the same as above.
%The defining polynomial $G(X)$ of  $(e_u)$
%is \\
%(1) If $q\equiv 1 \pmod 8$ and $p\equiv 5 \pmod 8$,
%$$
%G(X)=2\sum\limits_{j=0}^{q-1}X^{jp}+\sum\limits_{i=0}^{3} (\rho-i) D_i(X);
%$$
%(2) If $q\equiv 5 \pmod 8$ and $p\equiv 1 \pmod 4$,
%$$
%G(X)=2\sum\limits_{j=0}^{p-1}X^{jq}+2\sum\limits_{j=1}^{q-1}X^{jp}+(\rho+2) D_0(X)+(\rho+1)D_1(X)+\rho D_2(X)+(\rho+3)D_3(X).
%$$
%\end{theorem}

\begin{theorem}\label{def-poly-caseone}
If $q\equiv 1 \pmod 8$ and $p\equiv 5 \pmod 8$, the defining polynomial $G(X)$ of  $(e_u)$
is
$$
G(X)=2\sum\limits_{j=0}^{q-1}X^{jp}+\sum\limits_{i=0}^{3} (\rho-i) D_i(X),
$$
where $\rho=\sum\limits_{i=1}^{3} iD_i(\beta)$.
\end{theorem}

\begin{theorem}\label{def-poly-caseone-1}
If $q\equiv 5 \pmod 8$ and $p\equiv 1 \pmod 4$, the defining polynomial $G(X)$ of  $(e_u)$
is
$$
G(X)=2\sum\limits_{j=0}^{p-1}X^{jq}+2\sum\limits_{j=1}^{q-1}X^{jp}+\sum\limits_{i=0}^{3}(\rho+2-i) D_i(X),
$$
where $\rho=\sum\limits_{i=1}^{3} iD_i(\beta)$.
\end{theorem}
 To prove
Theorems \ref{def-poly-caseone} and \ref{def-poly-caseone-1}, we need some notations and auxiliary lemmas.

It is easy to see that
$$
uD_i:=\{uv \pmod{pq} : v\in D_i\}=D_{i+j}
$$
for $u\in D_j$.  We note that most of the calculations are performed in
the Galois ring $GR(4,4^{\ell})$ with characteristic four.

\begin{lemma}\label{lemma3}
Let $\gamma \in GR(4,4^{\ell})$ be a  primitive
$pq$-th root of unity,  then  we have

(1). $1+\gamma^{p}+\gamma^{2p}+\ldots+\gamma^{(q-1)p}=0$. %or $\gamma^{p}+\gamma^{2p}+\ldots+\gamma^{(q-1)p}=3$.

(2). $1+\gamma^{q}+\gamma^{2q}+\ldots+\gamma^{(p-1)q}=0$. %$\gamma^{q}+\gamma^{2q}+\ldots+\gamma^{(p-1)q}=3$ or %$1+\gamma^{q}+\gamma^{2q}+\ldots+\gamma^{(p-1)q}=0$.

(3). $\sum\limits_{z\in \mathbb{Z}_{pq}^*}\gamma^z=\sum\limits_{i=0}^3 D_i(\gamma)=1$.
\end{lemma}
It is easy to check these results, thus we omit the proof.

\begin{lemma}\label{lemma1}
Let $\gamma \in GR(4,4^{\ell})$ be a  primitive
$pq$-th root of unity. For $0\le i<4$, we have

(1). $D_i(\gamma^{kp})=0, ~~0\le k<q$.

(2). $D_i(\gamma^{kq})=3(q-1)/4, ~~1\le k<p$.
\end{lemma}
Proof. (1). Rewrite $s=\frac{q-1}{4}s_1+s_2$ with $0\le s_1 <\frac{p-1}{4} $ and $0\le s_2 <\frac
{q-1}{4} $, so from the definitions of $D_i$, $g$ and $h$, we have
%\begin{equation*}%\label{D_i-element}
%\begin{array}{lll}
%D_i & = & \{g^{(q-1)s_1+4s_2+i}h^j \pmod {pq} : 0\le s_1 <\frac{p-1}{4},0\le s_2 <\frac{q-1}{4}, 0\le j<4 \},
% \end{array}
%\end{equation*}
%and thus
$$D_i \pmod q = \{1,g^4,\ldots,g^{4((q-1)/4-1)}\} \pmod q :=D',$$
and each element in $D'$ appears $p-1$ times.
So for $0\le k< q$ we get by Lemma \ref{lemma3}(2)
$$D_i(\gamma^{kp})=(p-1)\sum\limits_{j\in D'}\gamma^{jkp}=0.$$
(2) Similarly, we have
 $$ D_i \pmod p  = \{1,\ldots,p-1\}=\mathbb{Z}_p^*,$$
and each element of $\mathbb{Z}_p^*$ appears $(q-1)/4$ times.
So we can get the desired result by Lemma \ref{lemma3}(2). ~ \hfill $\square$

\begin{lemma}\label{lemma2}
Let $0\le a<4$ and $w=g^{4x}h^j\in D_0$ for $0\le x< e/4$ and $0 \le j <4$.

(1). %$\sharp \{w\in D_0| g^a+w \equiv 0 \pmod p \}=\frac{q-1}{4}$.
 There are exactly $\frac{q-1}{4}$ many solutions $w$ satisfying $g^a+w \equiv 0 \pmod p$.

(2). There are exactly $p-1$ many solutions $w$ satisfying $g^a+w \equiv 0 \pmod q$ if $4 \mid (a+\frac{q-1}{2})$, and no solution otherwise .

(3). There is a solution $w$ satisfying both $g^a+w \equiv 0 \pmod p$ and $g^a+w \equiv 0 \pmod q$ iff
$4|(\frac{p-1}{2}-\frac{q-1}{2}-j)$ and $4 \mid (a+\frac{q-1}{2})$. Such solution is unique modulo $e/4$.
\end{lemma}
Proof.
(1) Since
$$
g^{4x} h^j \equiv -g^a=g^{(p-1)/2+a} \pmod p,
$$
then we have $4x+j\equiv (p-1)/2+a \pmod {p-1}$ and hence $4x+j=k(p-1)+(p-1)/2+a$ for all $0\le k<(q-1)/4$.

(2). Similarly, we have
$4x\equiv (q-1)/2+a \pmod {q-1}$ and hence $4x=k(q-1)+(q-1)/2+a$ for all $0\le k<(p-1)/4$ and $0\le j <4$.

For (3), we need to consider the equations
$$
\left\{
\begin{array}{ll}
4x\equiv (p-1)/2+a-j & \pmod {p-1},\\
4x\equiv (q-1)/2+a & \pmod {q-1}.
\end{array}
\right.
$$
By \cite[Lemma 5]{D97} and (2) of this lemma, we get the desired result. ~ \hfill $\square$

Below we  will calculate the inner product $\mathcal{C}_i(X)\cdot \mathcal{C}_j(X)^{\mathrm{T}}$ for $0\le i,j<4$, here  $\mathcal{C}_i(X)^{\mathrm{T}}$ is defined by  the transpose of $\mathcal{C}_i(X)$ and
$$
\mathcal{C}_i(X)=(D_{i}(X),D_{i+1}(X),D_{i+2}(X),D_{i+3}(X)).
$$

\begin{lemma}\label{inner-product}
Let $\beta \in GR(4,4^{\ell})$ be a  primitive
$pq$-th root of unity.  For any $0\le i,j<4$, we have
\begin{eqnarray*}
\mathcal{C}_i(\beta)\cdot \mathcal{C}_j(\beta)^{\mathrm{T}} + (q-1)/4
=\left\{
\begin{array}{ll}
1, & \mathrm{if}\,\ i=j, \\
0, & \mathrm{otherwise},
\end{array}
\right.
\end{eqnarray*}
if $q\equiv 1 \pmod 8$ and $p\equiv 5 \pmod 8$, and
\begin{eqnarray*}
\mathcal{C}_i(\beta)\cdot \mathcal{C}_j(\beta)^{\mathrm{T}} +(q-1)/4
=\left\{
\begin{array}{ll}
1, & \mathrm{if}\,\ i \equiv j+2 \pmod 4,\\
0, & \mathrm{otherwise},
\end{array}
\right.
\end{eqnarray*}
if  $q\equiv 5 \pmod 8$ and $p\equiv 1 \pmod 4$.
\end{lemma}
Proof. Since $D_i=g^{i} D_0$ for all  $0\le i<4$, we have
\begin{eqnarray*}
\mathcal{C}_i(\beta)\cdot \mathcal{C}_j(\beta)^{\mathrm{T}} & = & \sum\limits_{k=0}^{3}~\sum\limits_{u\in D_0}\beta^{ug^{i+k}}~\sum\limits_{v\in D_0}\beta^{vg^{j+k}}\\
          %& = & \sum\limits_{k=0}^{3}~\sum\limits_{u\in D_0}\beta^{ug^{i+k}}~\sum\limits_{w\in D_0}\beta^{uwg^{j+k}}~~ ~~(\mathrm{we~ use~ } v=uw)\\
          & = & \sum\limits_{k=0}^{3}~\sum\limits_{u\in D_0}~\sum\limits_{w\in D_0}\beta^{ug^{j+k}(g^{i-j}+w)}~~ ~~(\mathrm{here~ } v=uw)\\
          & = & \sum\limits_{w\in D_0}~\sum\limits_{k=0}^{3}~~\sum\limits_{z\in D_{j+k}}\gamma_w^z= %~~(\mathrm{we~ use~ } z=ug^{j+k}, \gamma_w=\beta^{g^{i-j}+w})\\
           \sum\limits_{w\in D_0}~\sum\limits_{k=0}^{3}D_{k}(\gamma_w),
\end{eqnarray*}
 and in the above penultimate  equation $ z=ug^{j+k}, \gamma_w=\beta^{g^{i-j}+w}$. Now we need to determine $\mathrm{ord}(\gamma_w)$, and we find that the possible values of $\mathrm{ord}(\gamma_w)$ are $1, p, q, pq$. Thus,
\begin{eqnarray*}
\mathcal{C}_i(\beta)\cdot \mathcal{C}_j(\beta)^{\mathrm{T}} & = & \Big(\sum\limits_{\stackrel{w\in D_0}{\mathrm{ord}(\gamma_w)=1}}
+\sum\limits_{\stackrel{w\in D_0}{\mathrm{ord}(\gamma_w)=p}}+\sum\limits_{\stackrel{w\in D_0}{\mathrm{ord}(\gamma_w)=q}}+\sum\limits_{\stackrel{w\in D_0}{\mathrm{ord}(\gamma_w)=pq}} \Big).
~\sum\limits_{k=0}^{3}D_{k}(\gamma_w)\\
\end{eqnarray*}

We first suppose that $q\equiv 1 \pmod 8$ and $p\equiv 5 \pmod 8$.  If $\mathrm{ord}(\gamma_w)= 1$, then $g^{i-j}+w \equiv 0 \pmod {pq}$. By Lemma \ref{lemma2}(3), there is unique $w(=g^{4s}h^t)\in D_0$ satisfying it iff $4|(\frac{p-1}{2}-\frac{q-1}{2}-t)$ and $4 \mid (\frac{q-1}{2}+i-j)$, that is, $t=2$  and  $i=j$.

If $\mathrm{ord}(\gamma_w)= p$, then $g^{i-j}+w \equiv 0 \pmod {q}$ but $g^{i-j}+w \not\equiv 0\pmod {p}$.  By Lemma \ref{lemma2}(2), there are $p-2$  many  or not any $w\in D_0$ satisfying it depending on whether $i=j$ or not.

Similarly, if $\mathrm{ord}(\gamma_w)= q$, %we find that $g^{i-j}+w \equiv 0 \pmod {p}$ but $g^{i-j}+w \not\equiv 0\pmod {q}$.
then by Lemma \ref{lemma2}(1), there are $\frac{q-1}{4}-1$ or $\frac{q-1}{4}$ many $w\in D_0$ satisfying it depending on whether $t=2$ and $i=j$ hold or not.

So the number of $w\in D_0$  satisfying  %$g^{i-j}+w \not\equiv 0\pmod {p}$ and $g^{i-j}+w \not\equiv 0\pmod {q}$, i.e.,
 $\mathrm{ord}(\gamma_w)= pq$, is $\frac{(p-1)(q-1)}{4}-(\frac{q-1}{4}-1)-(p-2)-1$ or $\frac{(p-1)(q-1)}{4}-\frac{q-1}{4}$ depending on whether $i=j$ or not.

From all the discussion above, we derive
\begin{eqnarray*}
\mathcal{C}_i(\beta)\cdot \mathcal{C}_j(\beta)^{\mathrm{T}} +(q-1)/4
=\left\{
\begin{array}{ll}
1, & \mathrm{if}\,\ i=j, \\
0, & \mathrm{otherwise}.
\end{array}
\right.
\end{eqnarray*}

For $q\equiv 5 \pmod 8$ and $p\equiv 1 \pmod 4$,  one can get the desired result by Lemma \ref{lemma2}(3) in a similar way.
%\begin{eqnarray*}
%\mathcal{C}_i(\beta)\cdot \mathcal{C}_j(\beta)^{\mathrm{T}} +(q-1)/4
%=\left\{
%\begin{array}{ll}
%1, & \mathrm{if}\,\ (i,j)\in\{(2,0),(3,1),(0,2),(1,3)\}, \\
%0, & \mathrm{otherwise}.
%\end{array}
%\right.
%\end{eqnarray*}
 ~\hfill $\square$

We are now ready to prove  Theorems \ref{def-poly-caseone} and \ref{def-poly-caseone-1}.

Proof of Theorem \ref{def-poly-caseone}. If  $q\equiv 1 \pmod 8$ and $p\equiv 5 \pmod 8$,
 we can check that the defining polynomial $G(X)$
of $(e_u)$ is
\begin{eqnarray*}
G(X)& =&
2 \sum\limits_{j=0}^{q-1}X^{jp}+ \sum\limits_{i=1}^3 i \Big( \mathcal{C}_k(\beta) \cdot \mathcal{C}_0(X)^{\mathrm{T}} +\frac{q-1}{4}\Big) \\
& =& 2\sum\limits_{j=0}^{q-1}X^{jp}+ \sum\limits_{i=1}^{3}
i \mathcal{C}_i(\beta)\cdot \mathcal{C}_0(X)^{\mathrm{T}}.
\end{eqnarray*}
In fact, for $u=kp$ with  $0\le k<q$, it follows from Lemma \ref{lemma1} that  $\mathcal{C}_0(\beta^{kp})=(0,0,0,0)$,  thus
$
G(\beta^{0})=2=e_{0}
$
and $
G(\beta^{kp})=0=e_{kp}
$
for $k\not=0$.

Similarly for $u= kq$ with $1\le k<p$, it follows from Lemma \ref{lemma3}(2) that $\mathcal{C}_0(\beta^{kq})=(\frac{3(q-1)}{4},\frac{3(q-1)}{4},\frac{3(q-1)}{4},\frac{3(q-1)}{4})$, so we have $G(\beta^{kq})=2=e_{kq}.$

For $u\in D_{k}$ with $0\le k<4$, Lemmas \ref{lemma3} and \ref{inner-product} leads to that
 $G(\beta^u)=k= e_u$.

Hence we get $e_u=G(\beta^u)$ for all $u\ge 0$.

With the above discussion and simple calculation, one can get the desired results from the definition of $\rho$ and Lemma \ref{lemma3}(3). ~\hfill $\square$

Proof of Theorem \ref{def-poly-caseone-1}. If  $q\equiv 5 \pmod 8$  and $p\equiv 1 \pmod 4$,  similar to the proof of Theorem \ref{def-poly-caseone}, one can check that the defining polynomial $G(X)$
of $(e_u)$ is
\begin{eqnarray*}
G(X)& =&2+
2\sum\limits_{j=0}^{p-1}X^{jq}+2\sum\limits_{j=0}^{q-1}X^{jp}+ \mathcal{C}_2(\beta)\cdot \mathcal{C}_3(X)^{\mathrm{T}}\\
&   & \qquad + 2\mathcal{C}_2(\beta)\cdot \mathcal{C}_2(X)^{\mathrm{T}}+3\mathcal{C}_2(\beta)\cdot \mathcal{C}_1(X)^{\mathrm{T}},
\end{eqnarray*}
thus we can get the desired results after simple calculation.
 ~\hfill $\square$

Theorems \ref{def-poly-caseone} and \ref{def-poly-caseone-1} are essential to the presentation of linear complexity of the sequences. Thus we have
\begin{theorem}\label{linearcomp}
If $q\equiv 1 \pmod 8$ and $p\equiv 5 \pmod 8$, the linear complexity of $(e_u)$
is
\[
L((e_u))=\left\{
\begin{array}{ll}
q+3(p-1)(q-1)/4, & \mathrm{if}\,\  2\in D_0,\\
pq-p+1, & \mathrm{if}\,\ 2\in D_2.
\end{array}
\right.
\]
\end{theorem}
\begin{theorem}\label{linearcomp-1}
If $q\equiv 5 \pmod 8$  and $p\equiv 1 \pmod 4$, the linear complexity of $(e_u)$ is
$$
L((e_u))=pq.
$$
\end{theorem}

%\begin{theorem}\label{linearcomp}
%The linear complexity $LC((e_u))$   of $(e_u)$
%is
%$$
%L((e_u))=pq-p+1~~~ \mathrm{or} ~~~  q+\frac{3(p-1)(q-1)}{4},
%$$
%if $q\equiv 1 \pmod 8$ and $p\equiv 5 \pmod 8$; and
%$$
%L((e_u))=pq~~~ \mathrm{or} ~~~  p+q-1+\frac{3(p-1)(q-1)}{4},
%$$
%if $q\equiv 5 \pmod 8$  and $p\equiv 1 \pmod 4$.
%\end{theorem}

Proof of Theorem \ref{linearcomp}. According to the work of Udaya and Siddiqi \cite[Theorem 4]{US}, we have  that
 the linear complexity $L((e_u))$ equals the number of nonzero coefficients of the defining polynomial of $(e_u)$. Then the result can be followed from Lemma \ref{rho-in-z4}.
~\hfill $\square$

Below, we let $\bar{b}$ denote the image of the element $b\in GR(4,4^r)$
under the natural epimorphism of the rings $GR(4,4^r)$
and $\overline{GR(4,4^r)} = GR(4,4^r)/2GR(4,4^r)$.

%Define $$E(x)=2\sum\limits_{i\in Q \cup R}^{3}x^i+ \sum\limits_{i=1}^{3} iD_i(x)$$ be the generating polynomial of the sequences $(e_u)$.
Define $$ E(x)=\sum\limits_{i=1}^{3} iD_i(x),$$ then it follows from  Lemma \ref{lemma3} and the relations of $D_l$ that  $E(\beta^k)=E(\beta)-l$ for  $k\in D_l$ with $l=0,1,2,3.$ Thus we have  $\overline{E(\beta^k)}= \overline{D_1(\beta)+D_3(\beta) }\in \mathbb{Z}_2$   iff $2|l$, i.e., $2\in D_0 \cup D_2$. Moreover, from our selection of $p$ and $q$, we have that $$2\in D_0 \cup D_2~\mathrm{iff}~ q\equiv 1 \pmod 8~\mathrm{ and~} p\equiv 5 \pmod 8,$$ since  $2$ is a quadratic residue (non-residue) modulo prime $q$ iff  $q\equiv \pm 1 \pmod 8$ ($q\equiv \pm 5 \pmod 8$).

\begin{lemma}\label{rho-in-z4}
$\rho \in \mathbb{Z}_4$ iff $2\in D_0$.
\end{lemma}
Proof. Let $H_i=D_i \cup D_{i+2}$ and $H_i(x) =\sum\limits_{j\in H_i}x^j$ for $i=0,1.$  We will prove the result with the following steps.

We first prove that if $q\equiv 1 \pmod 8$, then  $(H_0(\beta))^2=(0,0)H_0(\beta)+(0,1)H_1(\beta)$ with $(0,0)=|(H_0+1) \cap H_0 |$ and $(0,1)=|(H_0+1) \cap H_1 |$ are the generalized cyclotomic numbers of order $2$.

Note that $$(H_0(\beta))^2=\sum\limits_{u\in H_0}\sum\limits_{t\in H_0} \beta^{u(t+1)}.$$ We have  \cite[Proposition 9.8.6]{Ireland-Rosen} that $-1 \in D_0$ iff
 $q\equiv 1 \pmod 8$. One can see that $H_0$ contains $(q-1)/2-1$ elements such that $t+1 \equiv 0 \pmod p$, and $2(p-1)-1$ elements such that $t+1 \equiv 0 \pmod q$, respectively, for $t\not=-1$. It follows from Lemma \ref{lemma1} that
$$\sum\limits_{u\in H_0}\sum\limits_{t\in H_0, (t+1,pq)>1} \beta^{u(t+1)}=(\frac{q-3}{2})\cdot 0 +(2p-3)\frac{3(q-1)}{2}+\frac{(p-1)(q-1)}{2}=0.$$
Thus, we have $(H_0(\beta))^2=(0,0)H_0(\beta)+(0,1)H_1(\beta)$.

Second, we prove that $H_i(\beta)\in \mathbb{Z}_4$ for $i=0,1$ iff $2\in D_0\cup D_2.$
If $H_i(\beta)\in \mathbb{Z}_4$, then $\overline{H_i(\beta)}\in \mathbb{Z}_2$ and the discussion immediately above the Lemma leads to $2\in D_0\cup D_2.$ Meanwhile, if $2\in D_0\cup D_2,$ then $q\equiv 1 \pmod 8$. Denote $z=H_0(\beta)$, then by  Lemma \ref{lemma3} we have $z^2=(0,2)z+(0,1)(1-z)$, thus we have $z^2=z,$ i.e., $z\in \mathbb{Z}_2$   since $(0,0)=(q-5)(p-2)/4$ and  $(0,1)=(q-1)(p-2)/4$ \cite{D97}.

Finally, rewrite $E(x)=H_1(x)+2(D_2(x)+D_3(x))$, then if $E(\beta) \in \mathbb{Z}_4$,  we have $H_1(\beta)\in \mathbb{Z}_4$ and $ 2(D_2(\beta)+D_3(\beta))\in 2\mathbb{Z}_4$.  If  $2\in D_2$, then we have $ D_2(\beta^2)+D_3(\beta^2)=D_0(\beta)+D_1(\beta)$, thus we have $\overline{ D_2(\beta)+D_3(\beta)}^2=\overline{D_2(\beta)+D_3(\beta)}+1$, hence  $\overline{D_2(\beta)+D_3(\beta)} \not \in \mathbb{Z}_2.$ A contradiction.

If   $2\in D_0$,  we have $H_1(\beta)\in \mathbb{Z}_4$ and $\overline{D_2(\beta)+D_3(\beta)} \in \mathbb{Z}_2,$ thus $E(\beta)=\rho \in \mathbb{Z}_4$. ~~\hfill $\square$

 The proof of Theorem \ref{linearcomp-1} is similar to that of Theorem \ref{linearcomp}, we omit it here.

 Theorems \ref {linearcomp} and \ref {linearcomp-1} show that the sequence possess large linear complexity and they are ``good"  from the view point of cryptography.

\section{Final Remarks}

It is well known that the trace representation can be computed by applying the (discrete) Fourier transform \cite{M}.
 Trace functions over Galois rings \cite{US96,US}  is extensively applied to producing pseudorandom sequences efficiently
and analyzing their pseudorandom properties \cite{GG,KJKN,US} (see also references therein). The \emph{ trace function} $\mathrm{TR}_{s}^{r}(-)$ from $ GR(4,4^{r})$ to $ GR(4,4^{s})$ ($s|r$) is
defined by
$$
\mathrm{TR}_{s}^{r}(\alpha)=\Phi_s^0(\alpha) +\Phi_s(\alpha)+\ldots+\Phi_s^{r/s-1}(\alpha),
$$
where
$\Phi_s(\alpha)=\alpha_1^{2^s}+2\alpha_2^{2^s}$ is the \emph{Frobenius automorphism}  of $GR(4,4^{r})$ over $GR(4,4^{s})$ with order $r/s$.  For more details on trace functions over Galois rings, we refer the reader to \cite{W}.

Below we present the trace representation of  $(e_u)$ without proof since the proof is in a similar way as in \cite{Chen-Z4} with the fact that  $2\in D_0\cup D_2$ iff $q\equiv 1 \pmod 8$ and $p\equiv 5 \pmod 8$ and $2\in D_1\cup D_3$ iff $q\equiv 5 \pmod 8$ and $p\equiv 1 \pmod 4$.

%Proof. Note the fact that  $2$ is a quadratic residue (non-residue) modulo prime $q$ iff  $q\equiv \pm 1 \pmod 8$ ($q\equiv \pm 5 \pmod 8$), then with  the proof of  Lemma \ref{lemma1}(2) and (3), and the assumption $\gcd(p-1,q-1)=4$, the lemma can be followed.

\begin{theorem}\label{trace-caseone}
Let $\ell$ be the order of 2 modulo $pq$, $\ell_p$  the order of 2 modulo $p$ and $\ell_q$  the order of 2 modulo $q$. The trace representation of  $(e_u)$
is\\
(1) For $q\equiv 1 \pmod 8$ and $p\equiv 5 \pmod 8$,
$$
e_u=2+2\sum\limits_{i=0}^{\frac{q-1}{\ell_q}-1}\mathrm{TR}_{1}^{\ell_q}(\beta^{ug^ip})+\sum\limits_{i=0}^{3}(\rho-i)~ \sum\limits_{t=0}^{\frac{e}{4\ell/\epsilon}-1}\sum\limits_{j=0}^{3}\mathrm{TR}_{\epsilon}^{\ell}(\beta^{ug^{4t+i}h^j}),
$$
with $\epsilon=1$ if $2\in D_0$ and $\epsilon=2$ if $2\in D_2$.

(2) For  $q\equiv 5 \pmod 8$ and $p\equiv 1 \pmod 4$,
$$
e_u=2+2\sum\limits_{i=0}^{\frac{p-1}{\ell_p}-1}\mathrm{TR}_{1}^{\ell_p}(\beta^{ug^iq})
+2\sum\limits_{i=0}^{\frac{q-1}{\ell_q}-1}\mathrm{TR}_{1}^{\ell_q}(\beta^{ug^ip})
+\sum\limits_{i=0}^{3}(\rho+2-i)~ \sum\limits_{t=0}^{\frac{4e}{\ell}-1}\sum\limits_{j=0}^{3}\mathrm{TR}_{4}^{\ell}(\beta^{ug^{4t+i}h^j}).
$$
\end{theorem}

%
%\textbf{Fact II.}
%Let $\ell$ be the order of 2 modulo $pq$. Then $2|\ell$ if $2\in D_2$ and $4|\ell$ if $2\in D_1\cup D_3$.

%
\section*{Acknowledgements}
X.  Du  was partially supported by the National Natural Science
Foundation of China under grants No.  61462077 and 61702022. % (grants No. 61462077).
C. Wu. was partially supported by the National
Natural Science Foundation of China under grant No. 61373140 and
the Natural Science Foundation of Fujian Province under grant No. 2015J01662.

\end{document}